\documentclass[aps,prd,superscriptaddress,nofootinbib,12pt]{revtex4}

\usepackage{amsmath}
\usepackage{graphicx}
\usepackage{dcolumn}
\usepackage{bm}
\usepackage{amssymb}
\usepackage{latexsym}

\newcommand{\be}{\begin{equation}}
\newcommand{\ee}{\end{equation}}
\newcommand{\bq}{\begin{eqnarray}}
\newcommand{\eq}{\end{eqnarray}}
\begin{document}

\title{Couplings between holographic dark energy and dark matter}
\author{Yin-Zhe Ma}

\affiliation{Kavli Institute for Theoretical Physics China,
Institute of Theoretical Physics, Chinese Academy of Sciences,
P.O.Box 2735, Beijing 100080, China}
\affiliation{Institute of Astronomy, University of Cambridge,
Madingley Road, Cambridge CB3 0HA}
\author{Yan Gong}

\affiliation{National Astronomical Observatories,
Chinese Academy of Sciences, Beijing 100012, China}
\affiliation{Graduate School of Chinese Academy of Sciences,
Beijing 100049, China}
\author{Xuelei Chen}

\affiliation{National Astronomical Observatories,
Chinese Academy of Sciences, Beijing 100012, China}
\affiliation{Kavli Institute for Theoretical Physics China,
Institute of Theoretical
Physics, Chinese Academy of Sciences, P.O.Box 2735, Beijing 100080, China}

\begin{abstract}
We consider the interaction between dark matter and dark energy in
the framework of holographic dark energy, and propose a natural
and physically plausible form of interaction, in which the
interacting term is proportional to the product of the powers of
the dark matter and dark energy densities. We investigate the
cosmic evolution in such models. The impact of the coupling on the
dark matter and dark energy components may be asymmetric. While
the dark energy decouples from the dark matter at late time, just
as other components of the cosmic fluid become decoupled as the
universe expands, interestingly, the dark matter may actually
become coupled to the dark energy at late time. We shall call such
a phenomenon as \textit{ incoupling}. We use the latest type Ia
supernovae data from the SCP team, baryon acoustics oscillation
data from SDSS and 2dF surveys, and the position of the first peak
of the CMB angular power spectrum to constrain the model. We find
that the interaction term which is proportional to the first power
product of the dark energy and dark matter densities gives an
excellent fit to the current data.
\end{abstract}

\maketitle



\section{Introduction}

A number of astronomical observations, such as type Ia supernovae
(SNIa) \cite{SN}, the cosmic microwave background (CMB) anisotropy
\cite{CMB}, and the large scale structure (LSS) \cite{LSS}, have
shown that the expansion of our universe is accelerating. The
acceleration of the Universe strongly indicates the existence of a
mysterious exotic matter, named dark energy, which has negative
pressure and takes the largest proportion of the total density in
the Universe. The ultimate fate of our Universe is to be
determined by the dark energy, yet we know little about its
properties. Furthermore, while both the dark matter and the dark
energy is commonly thought to interact weakly with ordinary
matter, it is not immediately clear whether there is some
interaction between the two, or if the two are in some way linked.

In this paper we consider the dark matter-dark energy interaction
in the framework of the holographic dark energy (HDE) model \cite
{Cohen:1998zx,Horava:2000tb,Hsu:2004ri,Li:2004rb}. The basic ideas
of the HDE model are based on the holographic principle
\cite{holoprin}, which was inspired by the Bekenstein entropy
bound of black holes\cite{bh}. In this model, the dark energy is
related the quantum zero-point energy density $\rho _{\Lambda }$,
the total energy of the whole system with size $L$ should not
exceed the mass of a
black hole of the same size, thus we have $L^{3}\rho _{\Lambda }\leq LM_{%
\mathrm{pl}}^{2}$. The largest infrared cut-off $L$ is chosen by saturating
the inequality so that the HDE density is
\begin{equation}
\rho _{\mathrm{de}}=3c^{2}M_{\mathrm{pl}}^{2}L^{-2}~,  \label{holo-density}
\end{equation}%
where $c$ is a numerical constant, and $M_{\mathrm{pl}}\equiv
1/\sqrt{8\pi G} $ is the reduced Planck mass. If we take $L$ as
the size of the current Universe, for instance the Hubble scale
$H^{-1}$, then the dark energy density estimated with
Eq.(\ref{holo-density}) will be close to the observed value.
However, Hsu \cite{Hsu:2004ri} pointed out that this yields a
wrong equation of state for dark energy. Li \cite{Li:2004rb}
subsequently proposed that the future event horizon might be
chosen as the IR cut-off $L$, and viable HDE models can then be
constructed. As demonstrated in Ref. \cite{Li:2004rb}, the HDE
scenario may help to answer the {\it fine tuning problem} and the
{\it coincidence problem} of dark energy.

Various aspects of the HDE model and its extensions have been
investigated subsequently. For example, it has been generalized to
the case of non-flat Universe \cite{Huang04-2}, and it was studied
from the perspective of the anthropic principle \cite{Huang04-1}.
The holographic dark energy was also considered for driving the
inflation, and its resulting power spectrum index calculated
\cite{Enqvist04}. Ref.~\cite{Nojiri05} defined a new future event
horizon as the infrared cut-off $L$ and found that the coincidence
problem could be alleviated to some extent. Ref.~\cite{Hu06}
investigated the possible form of the cut-off scale $L$ for the
solution of coincidence problem. Inspired by the HDE model,
Ref.~\cite{Gao07} proposed the Ricci scalar curvature radius as
$L$. The model was reconstructed with a scalar field
\cite{ZhangMa05}. Observational constraints have also been studied
\cite{MGC07,Zhang05}. The growth of density perturbations in the
holographic dark energy models has been studied in \cite{Kim08}.

Interactions between the DE and DM components were considered for
the HDE model \cite{Pavon05,Wang05}. Ref.\cite{Pavon05} suggested
that for the $L=H^{-1}$ case, with interactions between the two
dark components, the dark energy could acquire an equation of
state in agreement with
observations, thus partly solving the no-acceleration problem noted in Ref.%
\cite{Li:2004rb} for interaction-free models. Ref.\cite{Wang05}
generalized the discussion in Ref.\cite{Pavon05} to the case of a
future event horizon. In that paper the proposed form of
interaction is $3H\rho _{tot}$, which could also lead to a
negative dark energy equation of state.

However, Ref.\cite{Miao08} pointed out that the interaction terms
such as $Q\sim 3H\rho _{tot}$, $3H\rho _{de}$, $3H\rho _{m}$ all
lead to unphysical results, because the dark matter density could
be driven to a negative value in these models.

Here we propose a form of interaction which we considered
\textit{natural} and \textit{physically plausible}. In general,
when two types of matter interact with each other, the interaction
should be proportional to the product of the powers of the number
density of both. This is most easily understood in the following
way: if one component does not exist, then there certainly would
be no interaction at all; with its amount increasing, the
interaction rate should increase. Thus, at least in the diffuse
limit, the interaction rate should be proportional to some powers
of the density (the relation could become more complicated when
the densities become high). For instance, this is the case for
ordinary two-body chemical reactions, 
namely
\begin{equation}
a^{-3}\frac{d(n_{1}a^{3})}{dt}\sim n_{1}n_{2}\left\langle \sigma
v\right\rangle ,  \label{Boltzmann}
\end{equation}%
i.e. the reaction rate is proportional to the product of number densities of
the two particle species.

However, besides the interaction in the usual sense (``scattering''),
one may also consider a more general
case where the dark energy is converted to dark matter or vice versa
by ``annihilation'' or ``decay''. In such a case, the ``interaction term'' or more
precisely the energy exchange term
would depend only on one of the densities and not on the other.

In general, we consider the interaction term to be proportional to
some powers of the density,
\begin{equation}
\dot{\rho}_{m}+3H\rho _{m}=\gamma \rho _{m}^{\alpha }\rho _{\Lambda }^{\beta
},  \label{dark matter}
\end{equation}%
\begin{equation}
\dot{\rho}_{\Lambda }+3H(1+w_{\Lambda })\rho _{\Lambda }=-\gamma \rho
_{m}^{\alpha }\rho _{\Lambda }^{\beta },  \label{dark energy}
\end{equation}%
in which $\rho _{m}$ and $\rho _{\Lambda }$ are the energy densities of the
matter and HDE respectively, and $\gamma $ is a parameter with mass
dimension $\mathrm{[density}^{\mathrm{\alpha +\beta -1}}\mathrm{\times time]}%
^{-1}$. The power index $\alpha,\beta$ is assumed to satisfy
$$\alpha,\beta \geq 0.$$
where if one of them is 0, we will have the case of ``decay'' (if the other is 1)
or ``annihilation'' (if the other is 2).
If $\gamma >0$, the interaction suggests that dark energy is converted
into dark matter, while $\gamma <0$ suggests the inverse process. This form
of interaction was considered in the study of coupled quintessence model
\cite{Mangano02}. However, not much research on this form of interaction
was carried out to understand how such a coupling term would affect
the evolution of the Universe.

In the next section, we present the evolution equations for the
interacting holographic dark energy model (details of our
derivation are given in the appendix). In Sec. 3, we study the
dynamical behavior of the model. In Sec.4
we explore the parameter space and find the best-fit and allowed region of $%
\alpha,\beta$. Finally, concluding remarks are given in Sec. 5.

\section{The model}
Now let us consider a spatially flat Friedmann-Robertson-Walker
(FRW) Universe filled with matter component $\rho _{m}$, HDE
component $\rho _{\Lambda }$ and the radiation $\rho _{r}$. For
simplicity we do not distinguish dark matter and ordinary matter
below but incorporate both in the matter component $\rho_m$. It is
certainly possible or even likely that dark energy only couples to
dark matter. However, in that case the baryons simply evolve as
$\rho_b \sim a^{-3}$ where $a$ is the scale factor, the baryon
component would then become dynamically irrelevant as the Universe
expands, so it only affects the details of the model but not its
qualitative behavior, hence we will neglect this effect in the
present paper. The Friedmann equation reads
\begin{equation}
3M_{pl}^{2}H^{2}=\rho _{m}+\rho _{\Lambda }+\rho _{r},  \label{Friedman1}
\end{equation}%
where $\rho _{r}$ is the energy density of the radiation which satisfies $%
\rho _{r}\sim a^{-4}$. Thus, Eqs. (\ref{holo-density}), (\ref{dark matter}),
(\ref{dark energy}), (\ref{Friedman1}) are the dynamic equations which
uniquely determine the evolution of $\rho _{m}(z),$ $\rho _{\Lambda }(z),$ $%
\rho _{r}(z),$ $w_{\Lambda }(z)$ and $H(z)$. In Eq.(\ref{holo-density}), $L$
could be Hubble horizon $H^{-1}$, particle horizon and future event horizon,
i.e.
\begin{equation}
L=R_{ph}(t)=a(t)\int_{0}^{a}\frac{da^{^{\prime }}}{H^{^{\prime }}a^{^{\prime
}2}},  \label{particle horizon}
\end{equation}%
or
\begin{equation}
L=R_{eh}(t)=a(t)\int_{a}^{\infty }\frac{da^{^{\prime }}}{H^{^{\prime
}}a^{^{\prime }2}}.  \label{event horizon}
\end{equation}%
We define the following dimensionless quantities: $\tilde{\rho}_{m}=\rho
_{m}/\rho _{0cr}$,$~\tilde{\rho}_{r}=\rho _{r}/\rho _{0cr}$, $~\tilde{\rho}%
_{\Lambda }=\rho _{\Lambda }/\rho _{0cr}$, where $\rho
_{0cr}=3H_{0}^{2}M_{pl}^{2}$ is the current critical density, and $%
E(z)=H(z)/H_{0}$, and introduce
\begin{equation}
\lambda =3^{\alpha +\beta -1}\gamma H_{0}^{2(\alpha +\beta
)-3}M_{pl}^{2(\alpha +\beta )-2}=\gamma \rho _{0cr}^{\alpha +\beta
-1}H_{0}^{-1},
\label{eq:lambda}
\end{equation}
The dimensionless $\lambda$ parameter determines the characteristic strength
of the interaction.

The equations of motion of the HDE model for the particle (event)
horizon cases are
\begin{eqnarray}
\frac{d\tilde{\rho}_{\Lambda }}{dz} &=&2\frac{\tilde{\rho}_{\Lambda }}{(1+z)}%
\left[1\pm \frac{\tilde{\rho}_{\Lambda }^{\frac{1}{2}}}{c\cdot E(z)}\right],\notag \\
\frac{dE(z)}{dz} &=&\frac{1}{2(1+z)}\left[3E(z)+\frac{(\tilde{\rho}_{r}-\tilde{%
\rho}_{\Lambda })}{E(z)} \pm \frac{2
\tilde{\rho}_{\Lambda }^{\frac{3}{2}}}{c\cdot E^{2}(z)}
-\frac{\lambda (E(z)^{2}-\tilde{\rho}_{r}-\tilde{\rho%
}_{\Lambda })^{\alpha }\tilde{\rho}_{\Lambda }^{\beta }}{E(z)^{2}}\right],  \notag \\
w_{\Lambda }(z) &=&-\frac{1}{3}\pm \frac{2}{3}\frac{\tilde{\rho}_{\Lambda }^{%
\frac{1}{2}}}{c\cdot E(x)}
-\frac{\lambda }{3}\frac{1}{E(x)}\left[E(x)^{2}-\tilde{%
\rho}_{\Lambda }-\tilde{\rho}_{r}\right]^{\alpha }\tilde{\rho}_{\Lambda }^{\beta
-1},  \label{event case}
\end{eqnarray}%
in which the upper(lower) sign represents the case of
particle(future event) horizon (c.f. the appendix for details of
derivation). It is easy to verify that if $L$ in
Eq.(\ref{holo-density}) equals the particle horizon $R_{ph}$ or
Hubble horizon $H^{-1}$, the model cannot fit the current
observational data, because current data prefer the dark energy
with equation of state near $-1$, while in the above two cases the
equation of state $w$ will be always greater than $-\frac{1}{3}$
(see also the analysis in Ref.\cite{Li:2004rb}). Therefore, in the
following we will focus on the interacting model with $L=R_{eh}$
(see Eq.(\ref{event case})).

\section{Dynamical behaviors}

In this section we study the dynamical behavior in models with
interaction between DE and DM, which may give rise to some
interesting behaviors. First we consider the case where
$\alpha,\beta$ are fixed to be 1, then consider the more general
cases.

As a typical example we consider the model with
$c=0.81,\lambda=-0.03$, which happens to be the $\alpha=\beta=1$
model which best-fits the current data ( for details of the
fitting procedure see next section).
First we plot the evolution of densities of different components in Fig.\ref%
{rho}. While one might think that the coupling between DE and DM
might help resolve the coincidence problem, we find that for these
particular parameter values the evolution of different densities
is rather similar to that of the standard $\mathrm{\Lambda CDM}$
model, i.e. the DM and DE densities could still differ by orders
of magnitude, so it does not seem to help much in this case.
Rather, as in other HDE models, the fine tuning problem could be
alleviated: the density of the dark energy is associated with the
scale of horizon rather than the Planck scale, due to the
holographic principle. We have also investigated cases with larger
coupling constant $\lambda$. We find that indeed, for the strong
coupling case ($\lambda>1$), the dark matter and dark energy could
be made to evolve with comparable density for an extended range of
redshifts, thus in some sense solving the ``coincidence problem''.
This is perhaps also true for any strongly coupled dark
matter-dark energy model. However, for strong coupling the fit to
observation is not good.

\begin{figure}[tb]
\centerline{\includegraphics[bb=0 0 437 288,width=4in]{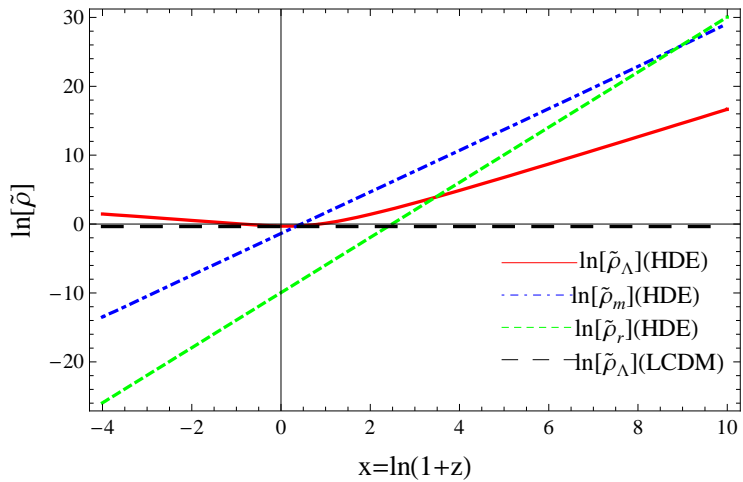}}
\caption{Evolution of densities in the HDE model. The black-dashed
line is the constant magnitude of dark energy in the LCDM
universe. The evolutions of matter and radiation in LCDM model are
very close to those (green-dashed and blue-dotted-dashed lines) in
the HDE model.} \label{rho}
\end{figure}

To investigate how the dynamics of the components are affected by coupling
with the other component, one may compare the expansion term $3H\rho_{i}$ ($%
i=m,\Lambda$) with the interaction term $\gamma \rho _{m}\rho
_{\Lambda }$. If the interaction term is greater than or
comparable with the expansion term of both of the components, the
two components are strongly coupled and evolve together. As
the Universe expands, typically the density of the component would drop as $%
a^{-3(1+w)}$, the interaction term often drops much more rapidly
than the expansion term. When the expansion term of a component is
much greater than the interaction term, that component evolves
almost freely, and we may say the two components are decoupled
\footnote{Note here that in the literatures the word
``decoupling'' may refer to a number of different phenomena. For
the decoupling of WIMP dark matter from the radiation-baryon
fluid, what is discussed above (the evolution of density) is
usually called ``thermal decoupling''. There is also ``kinetic
decoupling'', which refers to decoupling of temperature (see,
e.g., \cite{CKZ01}). Admittedly, this nomenclature is somewhat
confusing, since one might naively think that ``thermal
decoupling'' refers to decoupling of temperature, and ``kinetic
decoupling'' refers to decoupling of density evolution! Below we
shall refer to the decouplings discussed here as ``dynamical
decoupling'', the terms ``thermal decoupling'' and ``kinetic
decoupling'' are not used in this paper.}. The coupling and
decoupling effects between the two components could be asymmetric,
as each has its own expansion term. It is possible that for one
component the expansion term is greater, so it is decoupled from
the other component, but for the other component, its expansion
term is still smaller than or comparable with the interaction,
thus it is still coupled to the other component.

\begin{figure*}[tb]
\centerline{
\includegraphics[width=3.2in]{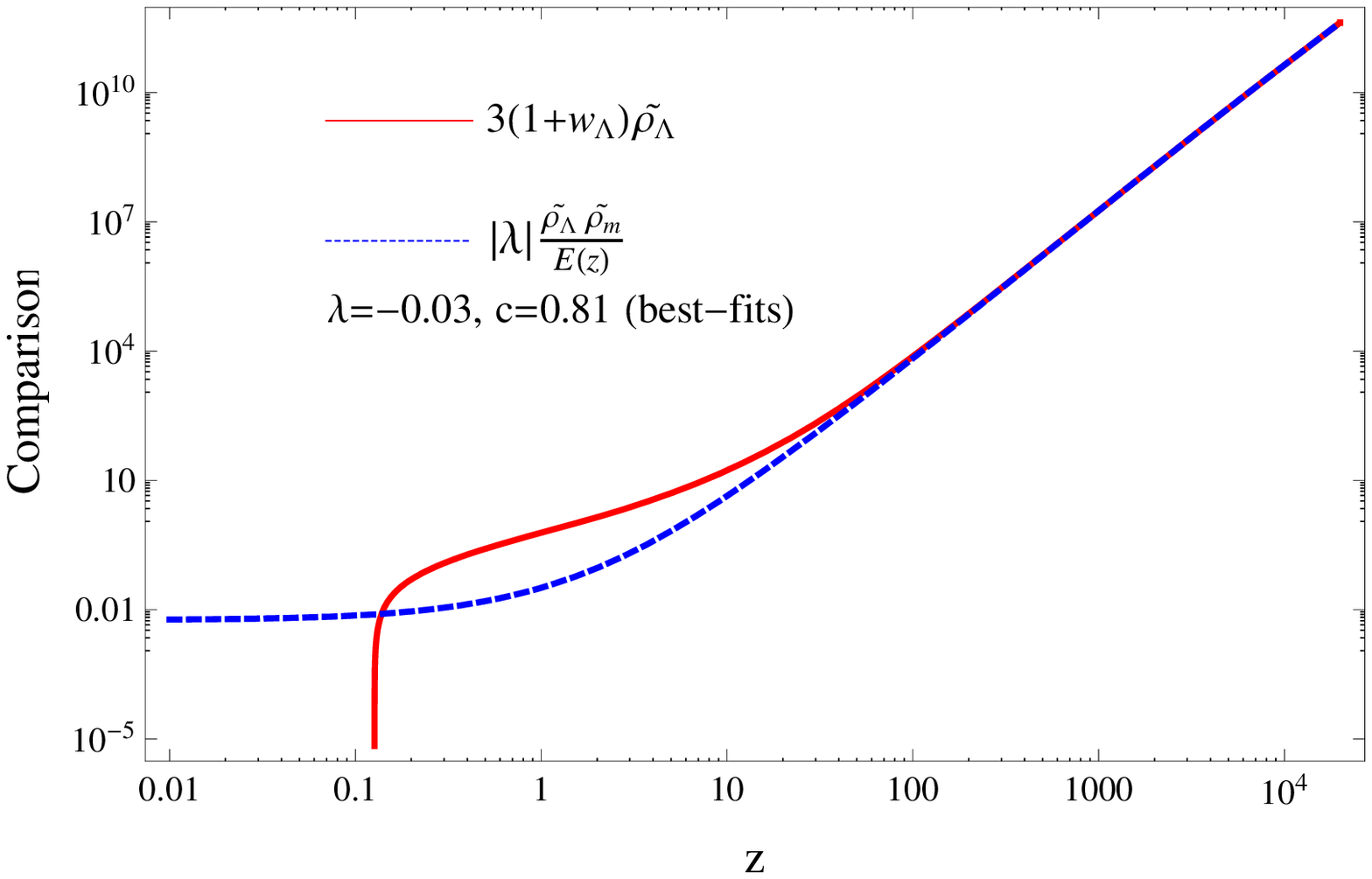}
\includegraphics[width=3.2in]{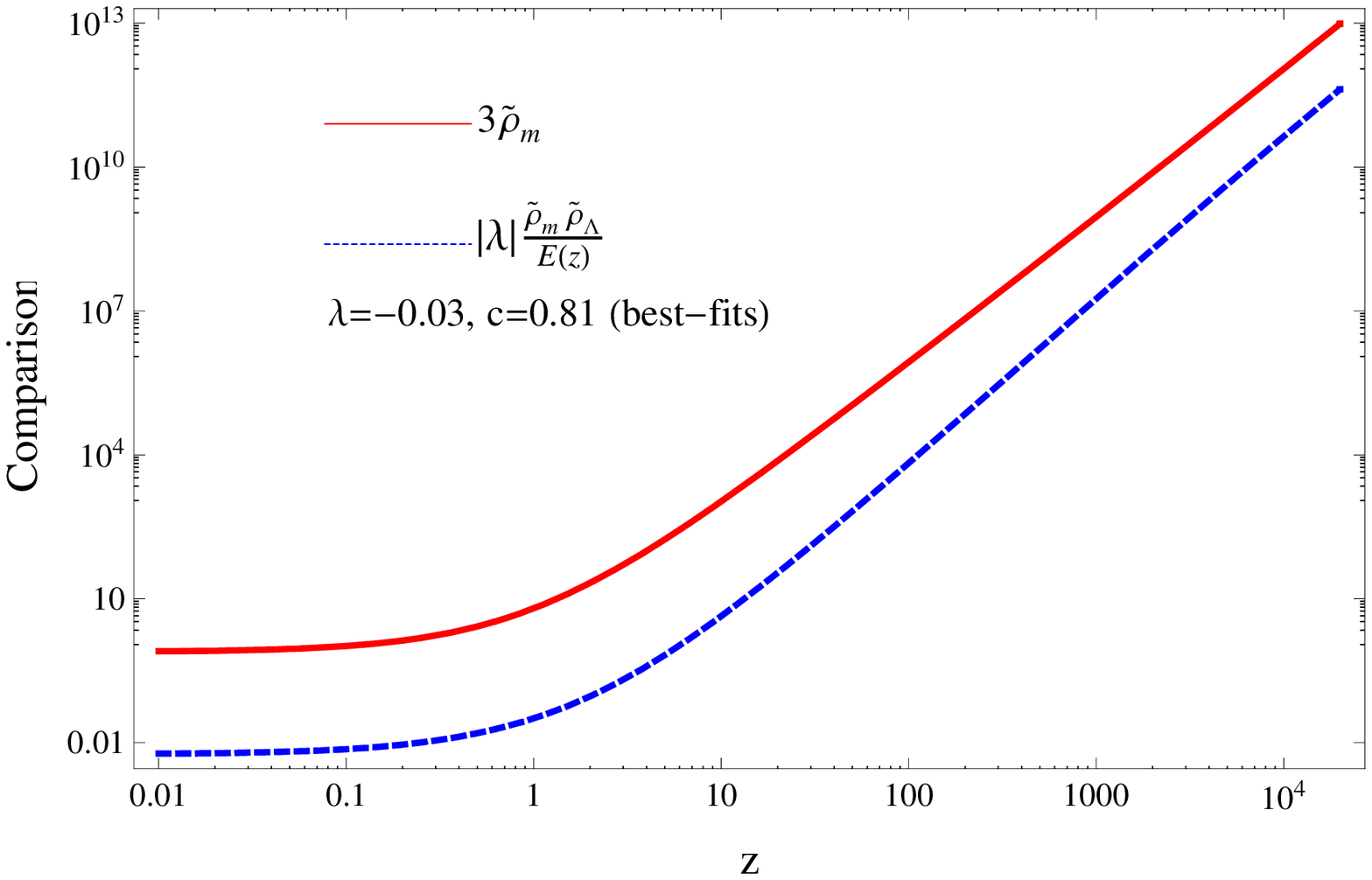}
} \caption{Comparisons of the expansion term (red solid line)
for dark energy (left panel) and  dark matter (right)
with the interaction term(blue dashed line) for the model with
$c=0.81, \lambda=-0.03$. }
\label{compbest}
\end{figure*}

In Fig.~\ref{compbest}, we compare the expansion term of the dark
energy and the dark matter with the interaction term as a function
of redshift. In the left panel, we plot the magnitude of the dark
energy expansion term $3(1+w) \tilde{\rho_{\Lambda}} $, and in the
right panel, we plot the magnitude of the dark matter expansion
term $3 \tilde{\rho}_m $, as red solid lines. The interaction term
$|\lambda| \frac{\tilde{\rho}_{m}\tilde{\rho}_{\Lambda }}{E(z)}$
is plotted in both panels for comparison with the expansion terms.
In the early Universe, the dark matter takes a greater proportion
of the total density, and for dark energy the interaction term is
comparable with the expansion term, so the dark energy is
moderately coupled to the dark matter. Later, as the density of
dark matter is diluted by expansion, the interaction term becomes
negligible compared with the expansion term, and the dark energy
is decoupled from dark matter. In the right panel of
Fig.~\ref{compbest}, we plot the dark matter expansion term
$3\tilde{\rho}_{m}$ and interaction term $\lambda
\frac{\tilde{\rho}_{m}\tilde{\rho}_{\Lambda }}{E(z)}$ as a
function of redshift $z$. We see that throughout the cosmic
history, for the dark matter the expansion term is always greater
than the interaction term, so the dark matter is at most weakly
coupled to dark energy. However, at lower redshifts, the
difference becomes larger, and the dark matter become further
decoupled from dark energy. Such a decoupling history is quite
typical for many interactions.

\begin{figure*}[tb]
\centerline{
\includegraphics[width=3.2in]{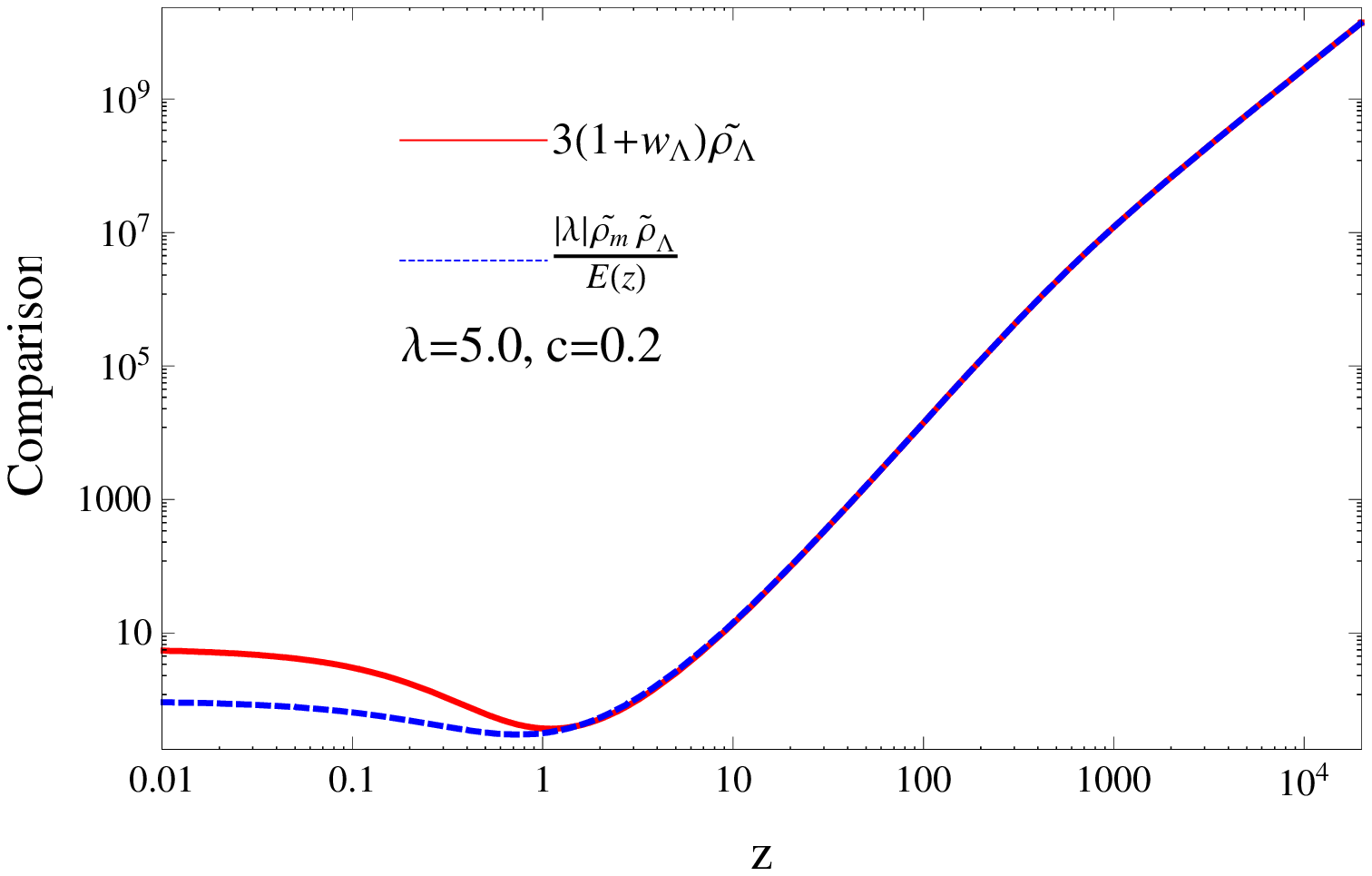}
\includegraphics[width=3.2in]{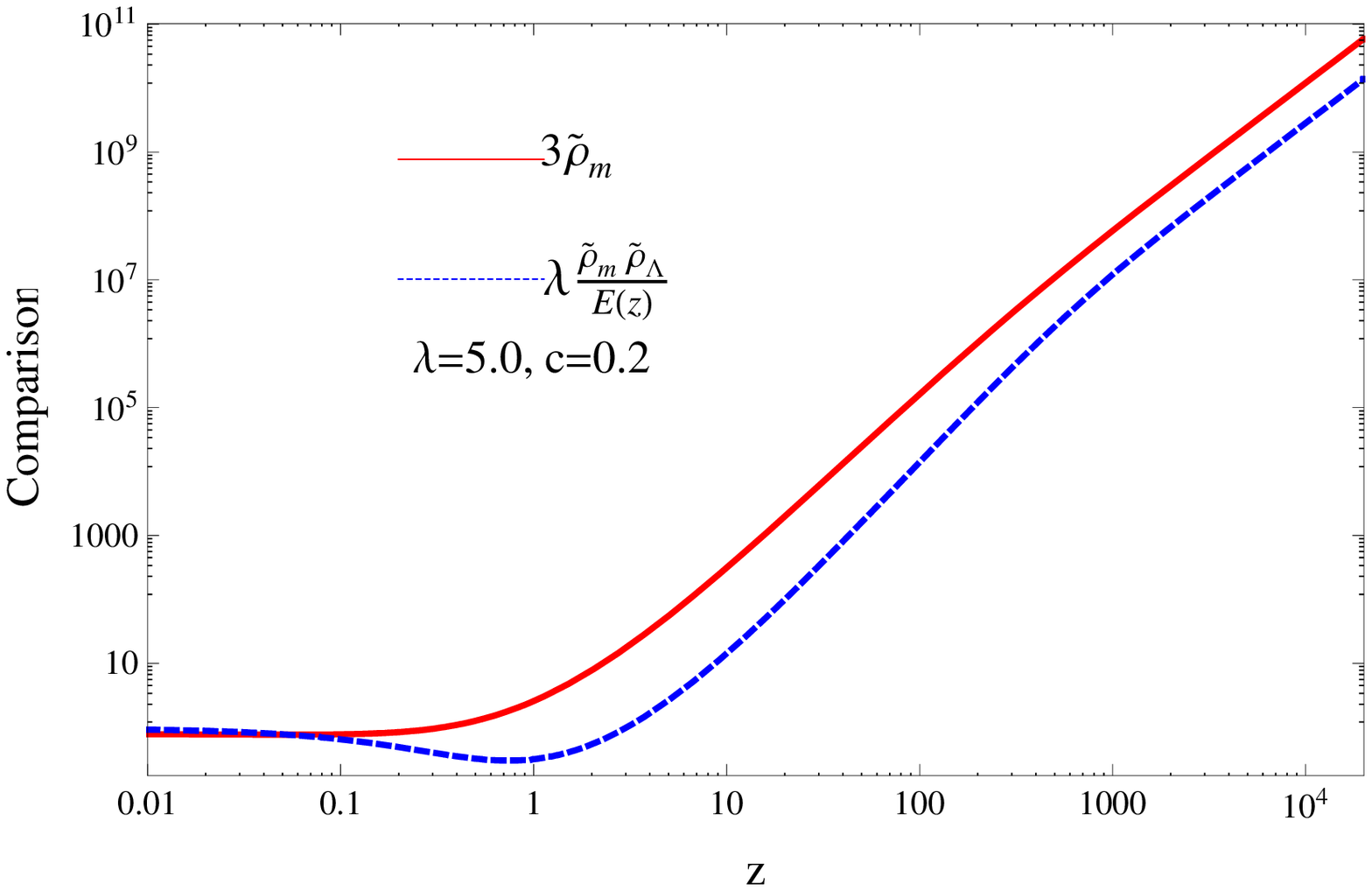}
}
\caption{Comparisons of the expansion term (red solid line)
for dark energy (left panel) and  dark matter (right)
with the interaction term(blue dashed line) for the model with
$c=0.2, \lambda=5$. }
\label{couple}
\end{figure*}

However, there could also be some quite unexpected and interesting
behavior in the presence of dark energy. In Fig.~\ref{couple}, we
make plots similar to Fig.~\ref{compbest} but for the model with
$c=0.2, \lambda=5$, and this model is chosen to illustrate such an
unusual behavior. In this model, because $w<-1$, the dark energy
density actually increases at later time. We see from the left
panel of Fig.~\ref{couple}, since initially the dark matter
density is greater, that the interaction term is comparable to the
expansion term for dark energy. However, later on the dark energy
becomes dominant, and the interaction term falls below the
expansion term of dark energy, so we may say that dark energy
becomes decoupled. However, more interestingly, from the right
panel of the figure, we see that initially the interaction term of
the dark matter is smaller than the expansion term, so the dark
matter is only weakly coupled to dark energy, but at later time,
the interaction term is comparable with the expansion term of dark
matter, so the DM becomes coupled to DE. This is the reverse of
the familiar decoupling process, which we shall call
\textit{incoupling}. This strange and interesting behavior could
only occur when a component is coupled to dark energy, whose
density does not decrease during the cosmic expansion. As the DE
keeps being converted into DM, the DM evolution is greatly
affected by DE. This model does not yield a good fit to the
observational data, but is only used to illustrate an interesting
dynamical behavior in the presence of $w<-1$ dark energy.

\begin{figure}[b]
\centerline{\includegraphics[bb=0 0 413 271,
width=3.6in]{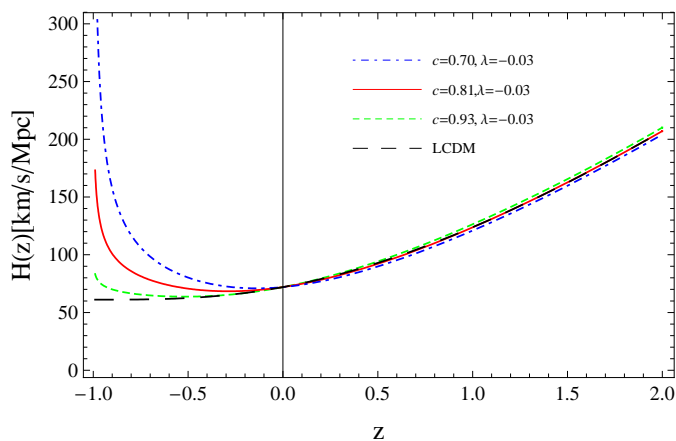}} \caption{Evolution of Hubble rate for
three choices of parameter $c$, three curves (red, blue, and green
one) are the best-fit, and $\pm 1 \protect\sigma_c$ C.L. values.
Here we set $\Omega _{m0}=0.26$,$H_{0}=72$km/s/Mpc. The
black-dashed line is the Hubble parameter in the LCDM universe.}
\label{Hubble}
\end{figure}

\begin{figure*}[tb]
\centerline{
\includegraphics[bb=0 0 395 260,width=3.3in]{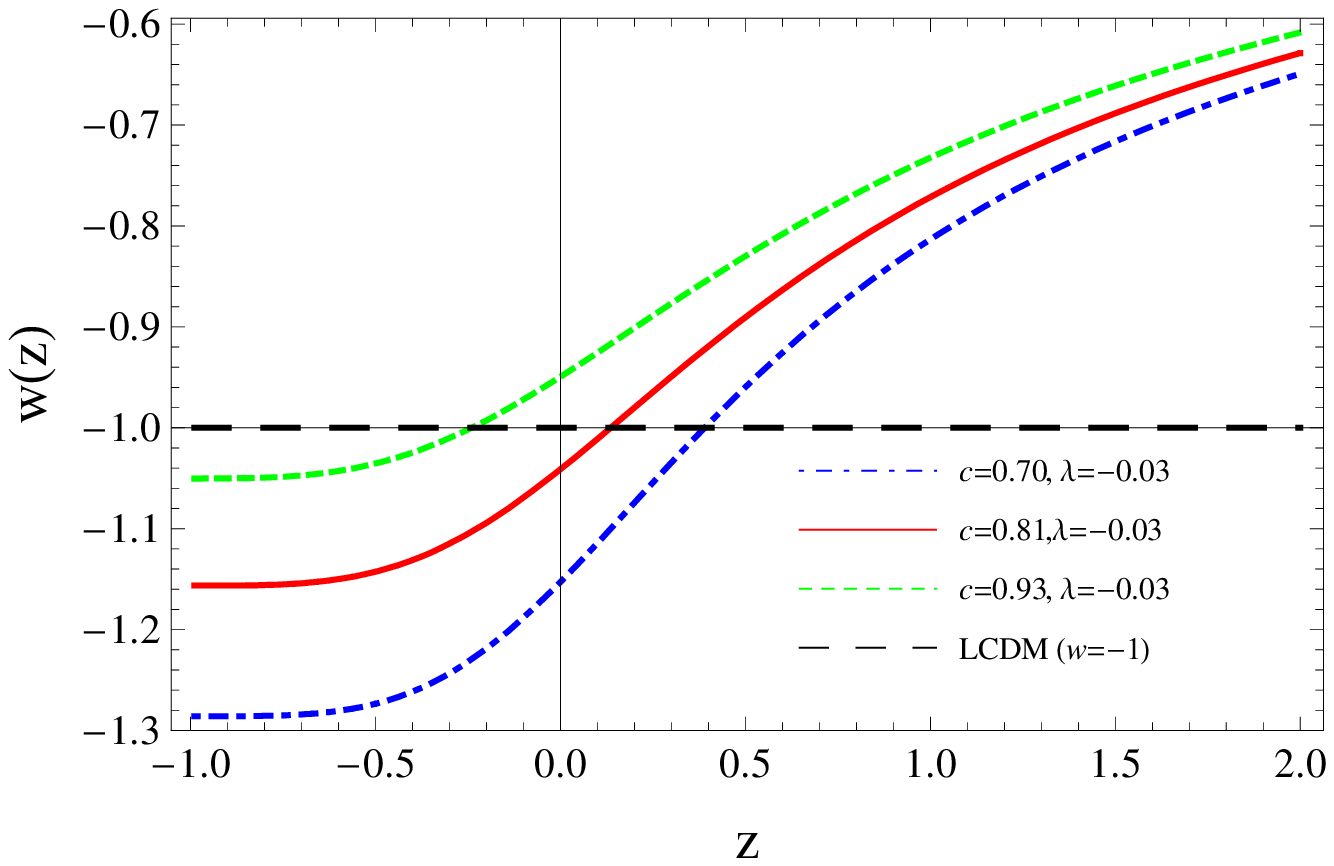}
\includegraphics[bb=0 0 392 252,width=3.3in]{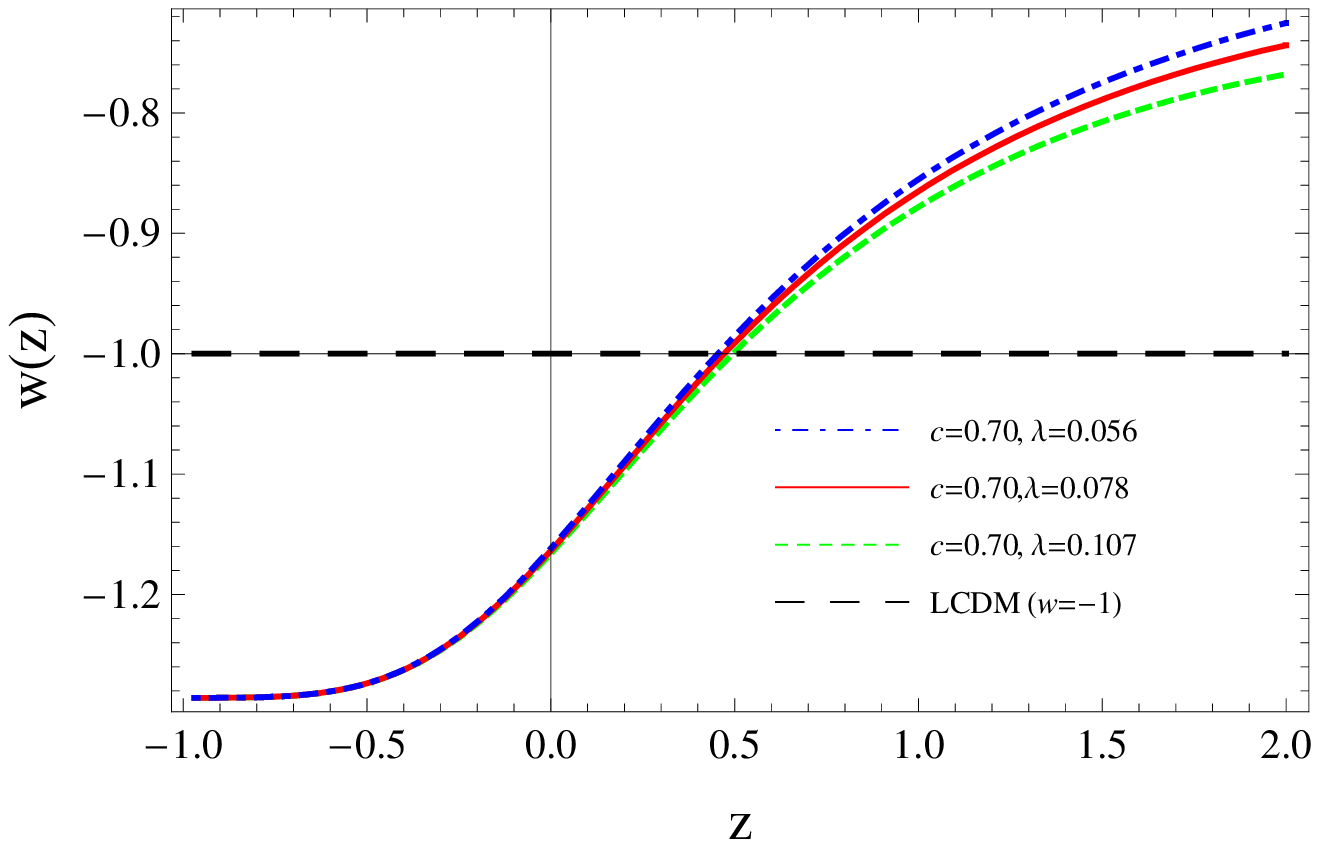}}
\caption{Equation of state with $\protect\alpha =\protect\beta =1$
for several parameters. The black-dashed line is the constant
equation of state in the LCDM universe.} \label{EoS1}
\end{figure*}

In Fig.\ref{Hubble}, we plot the evolution of $H$ for three
choices of $c$ parameter, i.e. the best fit $c_0=0.81$, and $c_0
\pm \sigma_c $, while keeping $\lambda=-0.03$. The Hubble parameter
for the $\Lambda$CDM model is also plotted for comparison.
We also plot the equation of state of the dark energy for these
models in Fig.~\ref{EoS1}. On the left panel, curves for different
$c$ but identical $\lambda$ are plotted; on the right panel,
curves for identical $c$ but different $\lambda$ are plotted. In
all of these models, the phantom divide $w=-1$ is either crossed,
or will be crossed in the near future. The dark energy density
would keep increasing, the interaction with DM would not slow it
down, as the dark energy has become decoupled to the DM.
Eventually a ``Big Rip'' would occur. The current value of the
equation of state $w_{0}$ for the best-fit model $c=0.81,
\lambda=-0.03$ is around $-1.04$, which is
within the $1\sigma $ CL region of $w_{0}$ in the WMAP 5-year results\cite%
{WMAP5}, suggesting that this model is consistent with the current
observations.

The value of $\lambda$ in these models does not significantly
affect the fate of the Universe, but it does greatly affect the
past evolution. This result is easy to understand; as the dark
energy dynamically decouples from dark matter, the interaction
become negligible, regardless the precise value of coupling
constant. The eventual fate of the Universe in this interacting
holographic dark energy model is qualitatively similar to those
without interactions, though the details of the evolution would be
different. Thus, as in the case of non-interacting holographic
models, for the fate of the Universe, $c<1$ leads to phantom-like
behavior and Big Rip\cite{Caldwell99}, while $c>1$ leads to
quintessence behavior. This is very different from the models
discussed in Ref.~\cite{MGC07}. For further discussions on the
fate of the Universe and Big Rip singularity in our model, see
Appendix B.

\begin{figure*}[tb]
\centerline{\includegraphics[bb=0 0 461 297, width=3.3in]{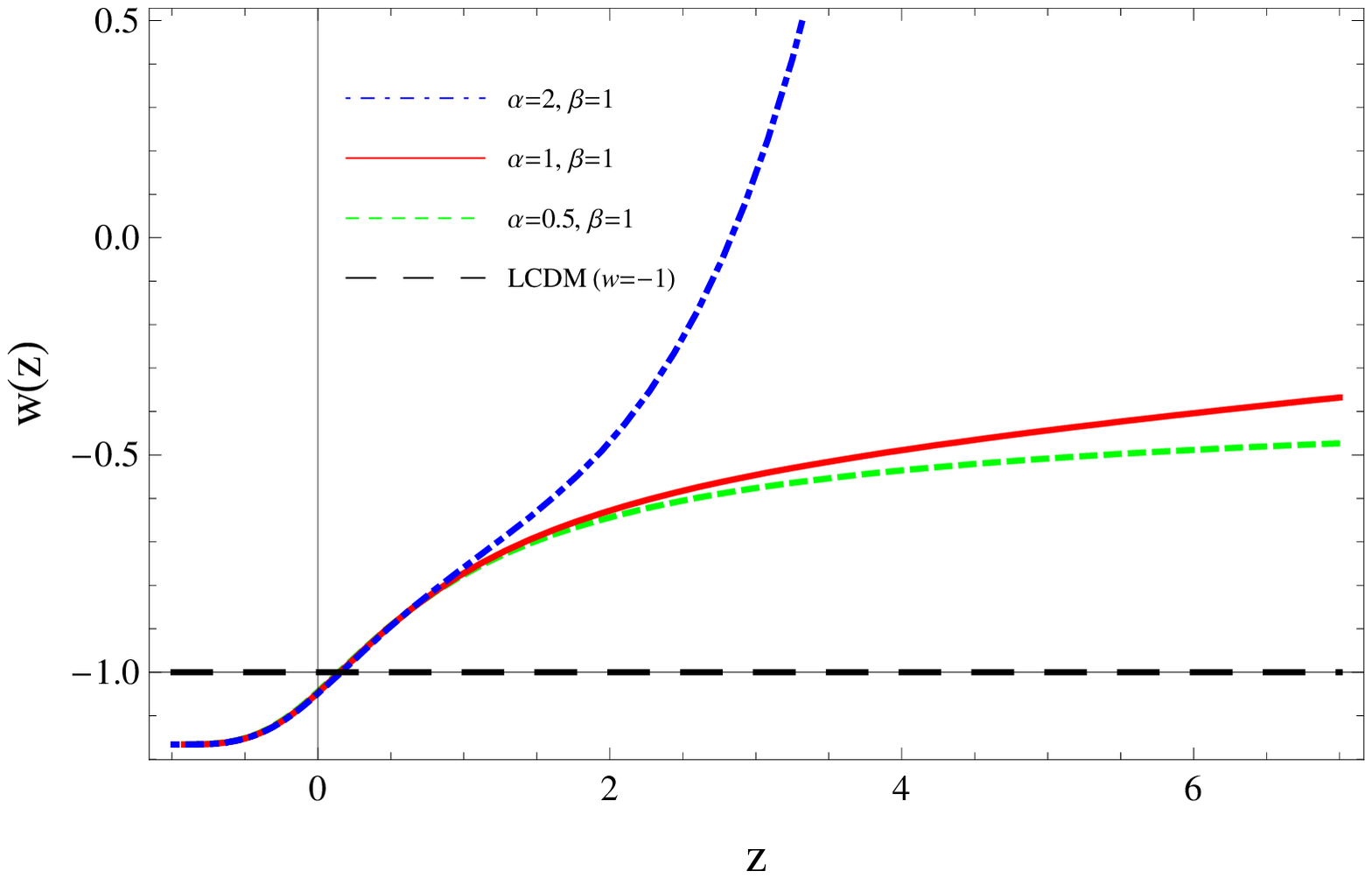}
\includegraphics[bb=0 0 430 280,width=3.3in]{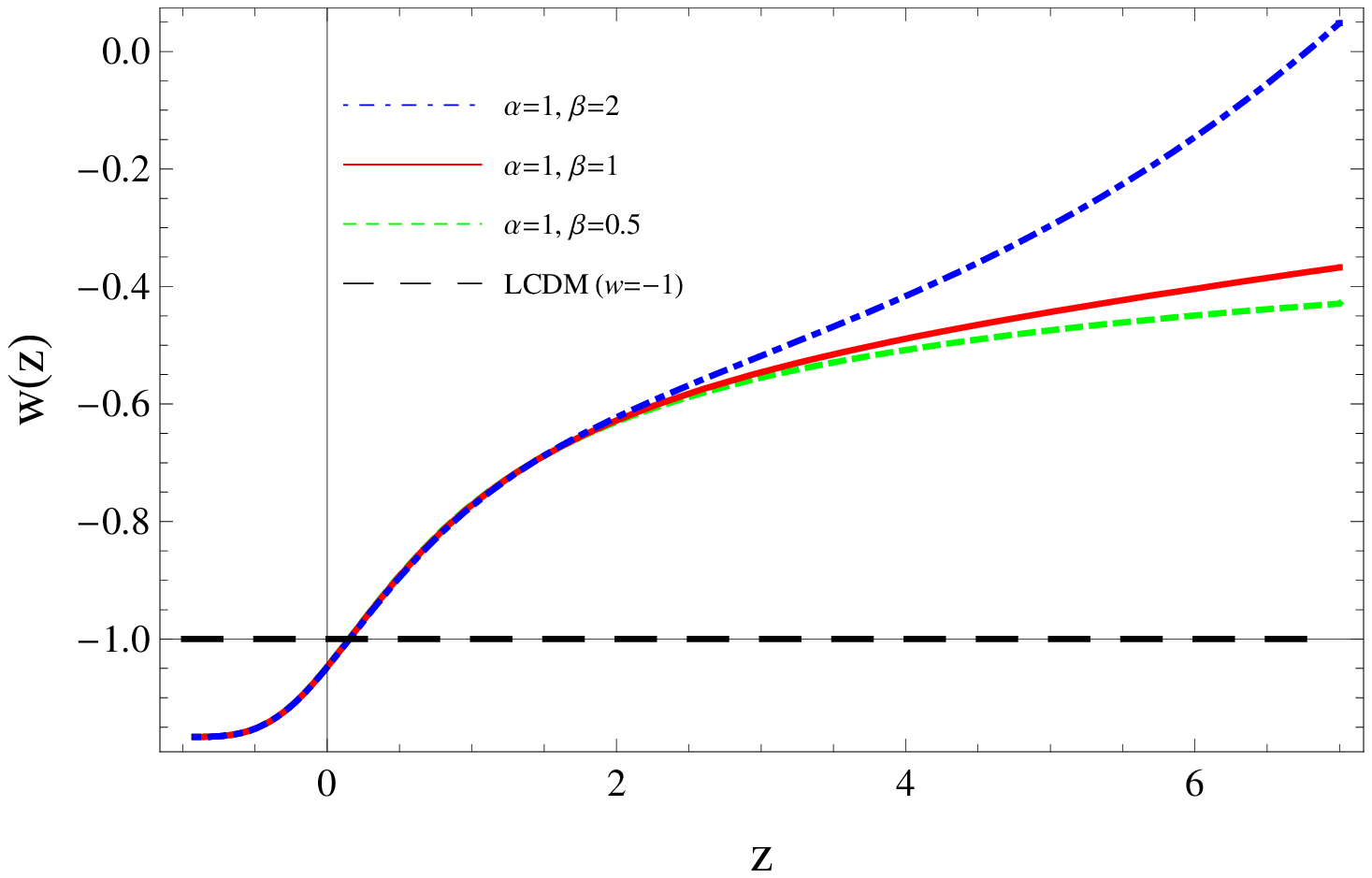}}
\caption{Redshift evolution of the equation of state, with the
best-fit values of $c$, $\lambda $ and $\Omega _{m0}$. The
black-dashed line is the constant equation of state in the LCDM
universe.} \label{EoS2}
\end{figure*}

We now consider the more general case of $\alpha ,\beta \neq 1$.
We plot the equation of state in Fig. \ref{EoS2}, varying $\alpha
$ in the left panel while keeping $\beta =1$, and varying $\beta $
in the right panel while
keeping $\alpha =1$. The other parameters are kept as before. For $\alpha <1$%
, the equation of state for dark energy is slightly more negative. For $%
\alpha >1$, at higher redshift $\rho _{m}^{\alpha }$ increases rapidly, and
the coupling term would become very large, the whole system would be
dominated by the evolution of matter. However, such a model would not really fit the low
redshift observation, so it would not be relevant to our purpose. For large $%
\beta $, we can also see that the equation of state of dark energy
$w>-1$ at high $z$, but the variation is not as rapid as in the
case of $\alpha $, and this is because the dark energy component
does not increase as much as the dark matter component. Still,
this makes a poor fit to the observational data.

Anyway, for the case $\alpha ,\beta \gg 1$, Eq.(\ref{EEoS}) is reduced to
\begin{equation}
w_{\Lambda }(x)\simeq -\frac{\lambda }{3}\frac{\tilde{\rho}_{\Lambda
}^{\beta -1}}{E(x)}(E(x)^{2}-\tilde{\rho}_{\Lambda }-\tilde{\rho}%
_{r})^{\alpha }.  \label{reduceEoS}
\end{equation}
Substituting this equation into Eq.(\ref{newrhoL}) and making use of Eq.(\ref%
{Fredman-new1}), we obtain the following approximate equation for $\tilde{%
\rho}_{\Lambda }$
\begin{equation}
\tilde{\rho}_{\Lambda }^{^{\prime }}\simeq 3\tilde{\rho}_{\Lambda },
\label{apprhoL}
\end{equation}
for which the solution will simply be
\begin{equation}
\tilde{\rho}_{\Lambda}(a)\simeq B /a^{3},
\end{equation}
where $B$ is a constant. This scaling behavior of dark energy
density (if $\alpha ,\beta \gg 1$) is very similar to that of the
Chaplygin gas model \cite{Bilic02}, which at the high redshift
also has the scaling behavior as $a^{-3}$ while at the low
redshift has the constant magnitude. However, in the present case
it does not provide a good fit to the observations.

\section{Fitting the model}

Now we fit the model parameters with the current observational
data. We use three kinds of observations; the first one is type Ia
supernovae (SNe Ia) which serves as the standard candle; the
second is the baryon acoustic oscillation (BAO) in large scale
structure of galaxy distribution, the third is the cosmic
microwave background anisotropies, the latter two provide standard
rulers at different redshifts. For the flat geometry we are
considering, $d_L=(1+z)^2 d_A= (1+z) r$, where $d_L, d_A, r$ are
the luminosity distance, angular diameter distance and comoving
coordinate distance, respectively, and $r$ is given by $r=
\frac{c}{H_0} \int_0^z \frac{dz'}{E(z')},$ which could be
calculated for each model by solving Eq.~(\ref{event case}) or
Eqs.~(\ref{motion3'})-(\ref{motion4''}).

\begin{figure}[tb]
\centerline{\includegraphics[width=3.4in]{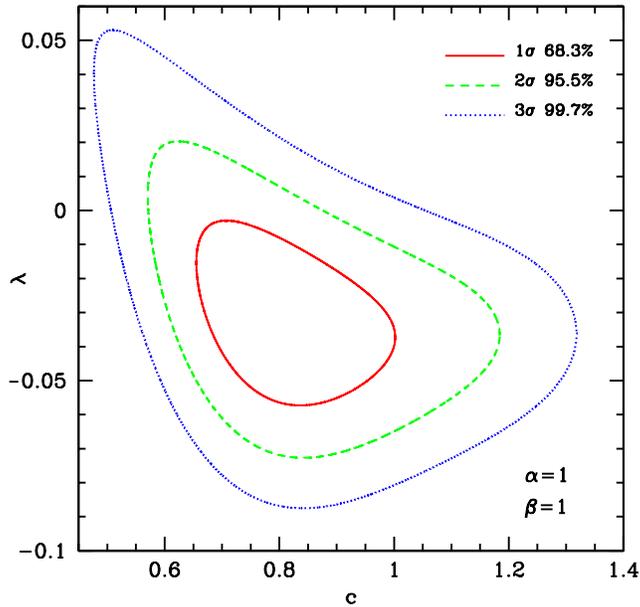}}
\caption{Contours of $\lambda$ versus $c$ for $\alpha=\beta=1$.}
\label{lambda0}
\end{figure}

For the SNe Ia, we use the data set published by Supernova
Cosmology Project (SCP) team recently \cite{SCP08}. This data set
contains 307 selected SNe Ia which includes several current widely
used SNe Ia data sets, such as those collected by the Hubble Space
Telescope (HST) \cite{Riess04, Riess06}, the SuperNova Legacy
Survey (SNLS) \cite{Astier05} and the Equation of State:
SupErNovae trace Cosmic Expansion (ESSENCE) surveys\cite{Wood07}.
Using the same analysis procedure and improved selection approach,
all of the sub data sets are analyzed to obtain a consistent and
high-quality ``Union'\ SNe Ia data set, which provides tighter and
more reliable constraints for our models. For the BAO data, we use
the data given in Percival \emph{et al}.
\cite{Percival07} and WMAP team \cite{WMAP5}. We use the quantity $%
r_{s}/D_{v}$ which is determined by the Sloan Digital Sky Survey (SDSS) ($%
z=0.35$) and the Two Degree Field Galaxy Redshift Survey (2dFGRS)
($z=0.2$) in our constraints, where $r_{s}$ is the comoving sound
horizon size at the decoupling epoch, and $D_{v}$ is the effective
distance defined by Eisenstein \emph{et al}. \cite{Eisenstein05}.
Since this quantity is just a ratio of two ``standard rulers'', it
could give a reliable measurement of the expansion history of the
Universe. For the CMB, we use the position of the
first peak of the Cosmic Microwave Background (CMB) angular power spectrum $%
\ell_1$. The details of this technique were described in our
earlier paper Ref. \cite{Gong08}.

We use the Markov Chain Monte Carlo (MCMC) method to constrain the
parameters with the combined data set of SN Ia+BAO+CMB (for details of our
implementation of this technique, see Ref.~\cite{Gong07}).
In the present model, the relevant
parameters are the matter density $\Omega_{m0}$, Hubble constant $H_0$, and
the holographic dark energy parameter $c,\lambda$. We marginalize over $H_0$
and $\Omega_{m0}$. For $\alpha, \beta$, we limit their range to $%
0<\alpha,\beta<2$. As discussed in the last section, greater values of $%
\alpha, \beta$ would not yield a good fit to observation.

\begin{figure*}[tb]
\centerline{\includegraphics[width=3.4in]{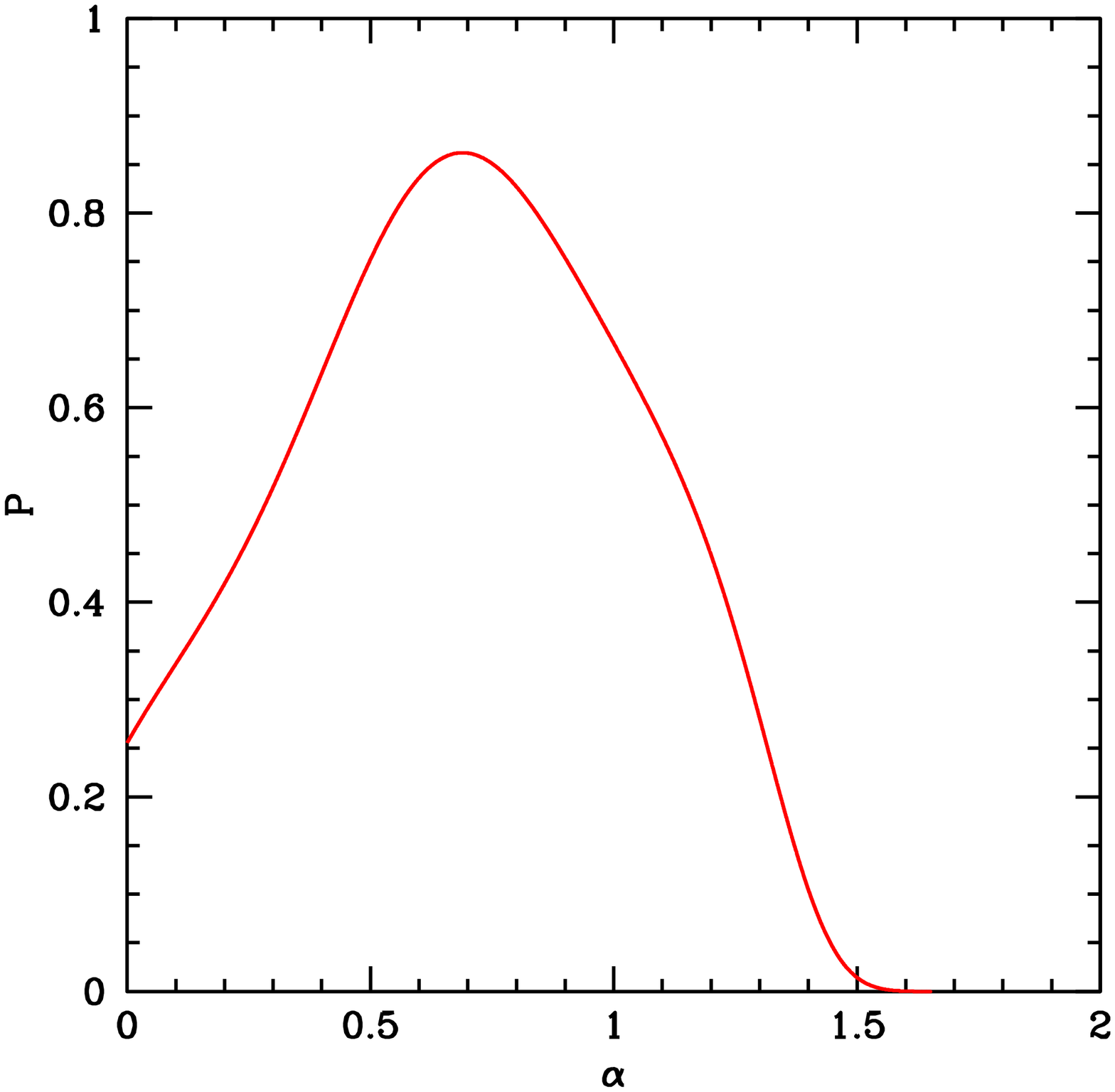}
\includegraphics[width=3.4in]{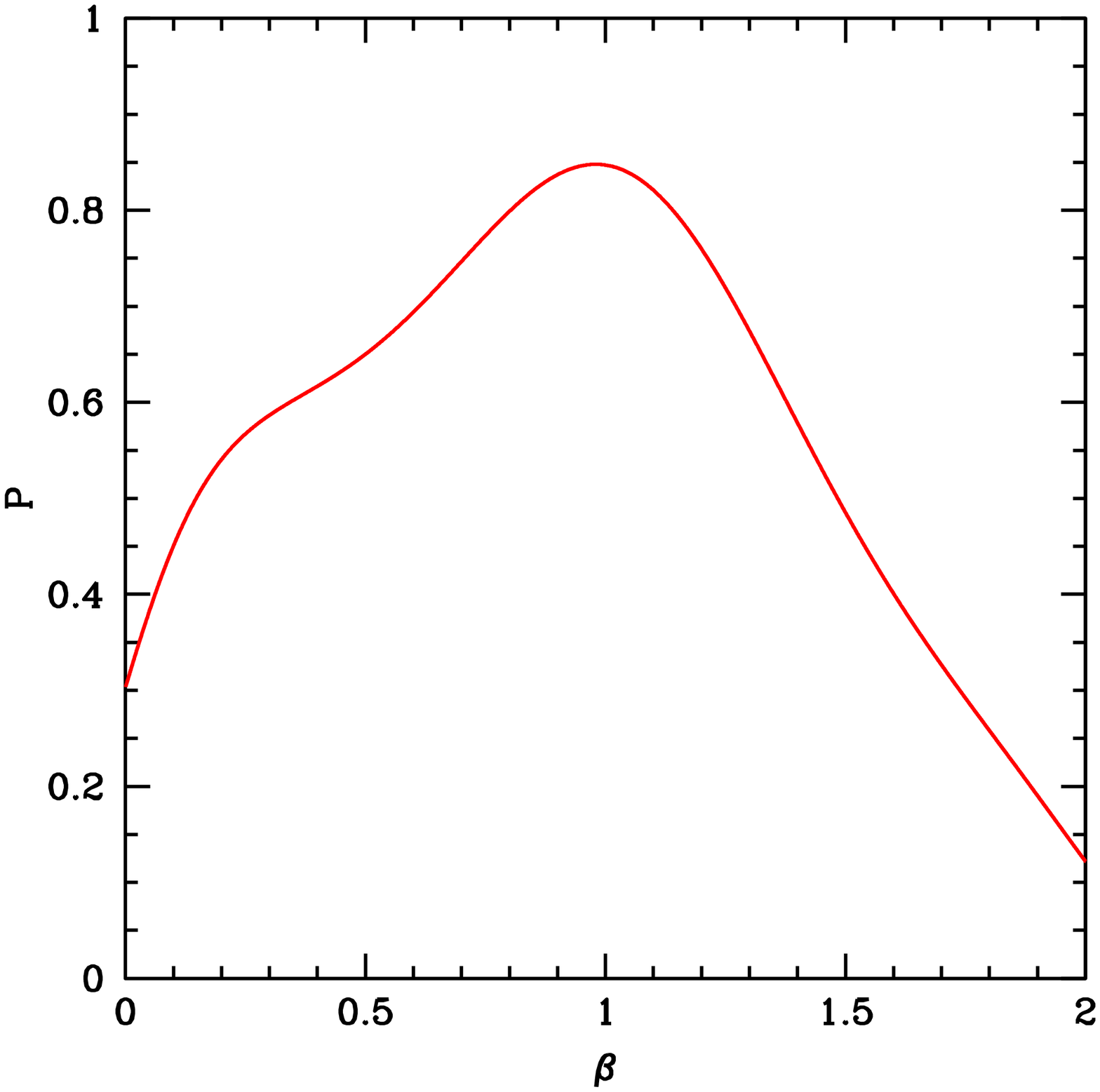}}
\caption{Probability distribution function of the power index parameters $%
\protect\alpha $ and $\protect\beta$.}
\label{alpha}
\end{figure*}

For the fixed $\alpha=\beta=1$ case, as can be seen in Fig.
\ref{lambda0}, the best-fit value and error on $c,\lambda$ are
\begin{equation}
c=0.81^{+0.12}_{-0.11}, \quad
\lambda=-0.03\pm 0.02.
\end{equation}
The behavior of the best fit model has been discussed as the example given in
the previous section. The models within the $1\sigma$ contour also exhibit
similar behavior.

\begin{figure*}[htb]
\centerline{\includegraphics[width=3.4in]{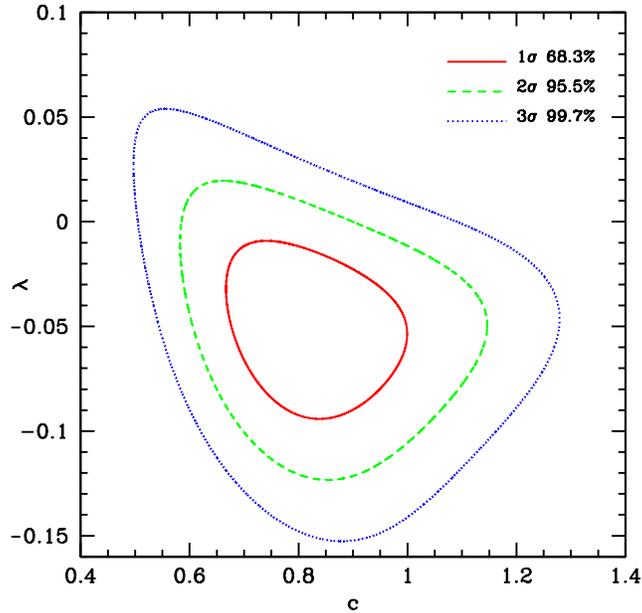}}
\caption{Contours of $\lambda$ versus $c$ for free $\alpha$ and $\beta$.}
\label{lambda}
\end{figure*}

Now we let $\alpha,\beta$ be free parameters, the PDF of $\alpha$
and $\beta$ are shown in Fig. \ref{alpha}. The best-fit values for
$\alpha,\beta$ are $\alpha=0.69^{+0.45}_{-0.38}$ and
$\beta=0.98^{+0.49}_{-0.81}$. If one demand $\alpha,\beta$ to be
integers for physical reasons, then $\alpha=1,\beta=1$ is still
the best fit. Alternatively, a non-integer power may rise if the
interaction is complicated. The best fit for $c$ and $\lambda$ are
$c=0.81^{+0.13}_{-0.11}$ and $\lambda=-0.04^{+0.03}_{-0.04}$, as
shown in Fig. \ref{lambda}, it is not too different from the
$\alpha=\beta=1$ case.

\section{Conclusion}

The holographic dark energy (HDE) model is a dynamical model of dark energy,
motivated by the holographic principle, in which the event horizon of
the space-time plays an important role in the dynamics.
The microphysical basis of the model is not well understood, but the
dynamics of cosmic expansion could be studied. Observationally, the
HDE model is distinguishable from the $\Lambda$CDM, as its equation of
state varies with time. The model could also be distinguished from other
dynamical dark energy models. For example, in the HDE model the
dark energy behaves as the Chaplygin gas at high redshift if
$\alpha,\beta \gg 1$ (see the discussion at the end of Sec.3),
but for the low redshift region, the equation of state of
the Chaplypin gas approaches a constant, while for this model
it continues to evolve. In properly constructed HDE models such as
the ones discussed in this paper, the dark energy
component is subdominant at high redshifts compared with matter or
radiation, so it would not significantly affect
the CMB or large scale structure power spectrum, though at low redshift
it could affect the growth of structure in detail. We plan to investigate
the details of the structure growth in subsequent studies.

In the framework of the holographic dark energy model, we proposed
a form of interaction between dark energy and dark matter which we
considered to be natural and physically plausible. That is, the
interaction rate is proportional to the product of powers of the
dark energy and dark matter densities $\rho_m^\alpha
\rho_{\Lambda}^\beta$. For this type of interaction, we find that
for the weak coupling case (the dimensionless coupling constant
$\lambda<1$, see Eq. (\ref{eq:lambda})), the ultimate fate of the
Universe is similar to the case of non-interacting holographic
dark energy, namely for $c<1$ the equation of state of the dark
energy will cross $w=-1$, and behave like a phantom in the future,
and the Universe ends in a Big Rip. We also discovered the
interesting phenomenon of \textit{incoupling}, which occurs only
when the dark matter is coupled to dark energy. This is the
opposite of the familiar process of {\it decoupling}, the dark
matter becomes increasingly coupled to the dark energy as Universe
expands. We then use the currently available observational data to
constrain the parameters of this model. The data set we used
include 307 selected SNe Ia published by the SCP team
\cite{SCP08}, baryon acoustic oscillation from SDSS and 2dFGRS
\cite{Percival07}, as well as the position of the first peak of
the CMB angular power spectrum \cite{WMAP5}. We find that the
best-fit form of interaction is $\alpha=0.69,\beta=0.98$. If one
requires $\alpha,\beta$ to be integers for reasons of fundamental
physics, then the most favored interaction rate is proportional to
the product of the density of dark matter and dark energy.

 \begin{acknowledgments}
Our MCMC chain computation was performed on the Supercomputing
Center of the Chinese Academy of Sciences and the Shanghai
Supercomputing Center. This work is supported by the Chinese
Academy of Sciences under grant KJCX3-SYW-N2, by the Ministry of
Science and Technology National Basic Science program (project 973)
under grant No.2007CB815401, and by the National Science Foundation of China
under the Distinguished Young Scholar
Grant 10525314, and the Key project Grants No.10503010.

\end{acknowledgments}

\appendix

\section{dynamical equations}

\begin{figure*}[tb]
\centerline{\includegraphics[width=3.9in]{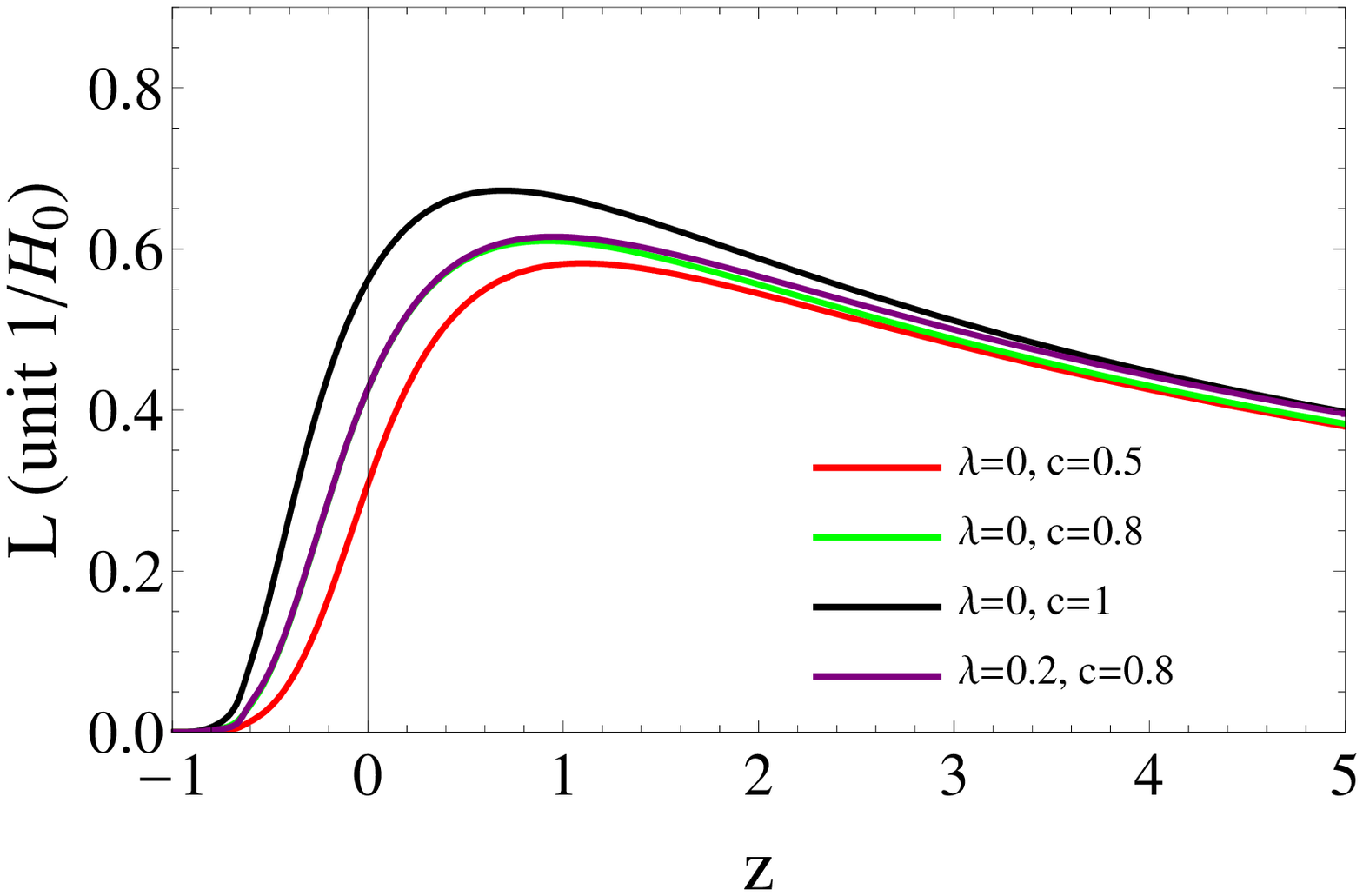}
\includegraphics[width=3.6in]{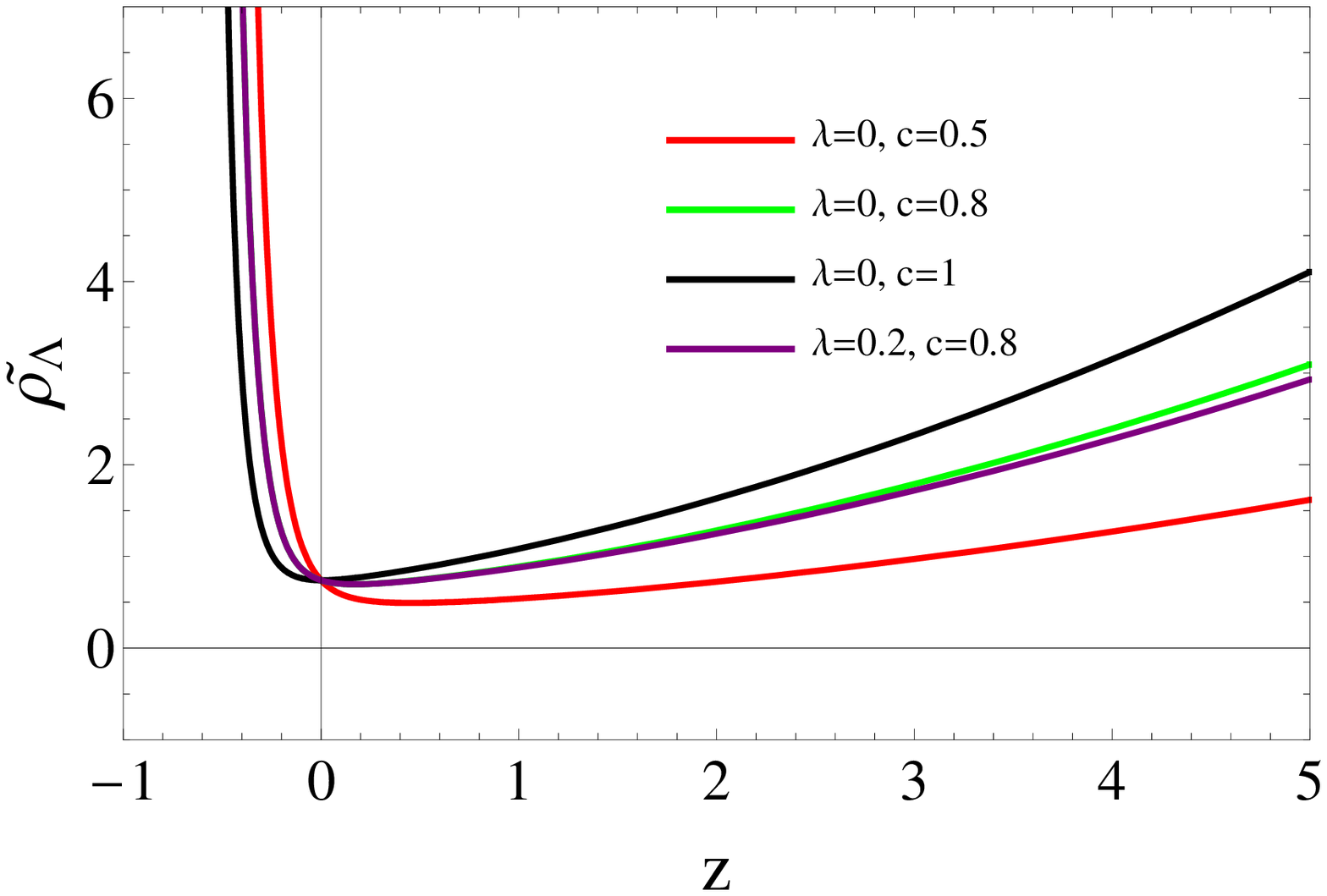}}
\caption{Left: the evolution of event horizon size; right: the
evolution of dark energy density.} \label{diverge}
\end{figure*}

We give the detailed derivation of the dynamical equations of the
interacting holographic dark energy model here. Using the dimensionless
notations given in Sec.2, we have
\begin{equation}
E(x)^{2}=\tilde{\rho}_{m}+\tilde{\rho}_{\Lambda }+\tilde{\rho}_{r},
\label{Fredman-new1}
\end{equation}%
\begin{equation}
\tilde{\rho}_{r}=\Omega _{r0}e^{4x},  \label{rad density}
\end{equation}%
\begin{equation}
\tilde{\rho}_{m}^{^{\prime }}-3\tilde{\rho}_{m}=-\lambda \frac{\tilde{\rho}%
_{m}^{\alpha }\tilde{\rho}_{\Lambda }^{\beta }}{E(x)},  \label{new rhom}
\end{equation}%
\begin{equation}
\tilde{\rho}_{\Lambda }^{^{\prime }}-3(1+w_{\Lambda })\tilde{\rho}_{\Lambda
}=\lambda \frac{\tilde{\rho}_{m}^{\alpha }\tilde{\rho}_{\Lambda }^{\beta }}{%
E(x)},  \label{newrhoL}
\end{equation}%
\begin{equation}
\tilde{\rho}_{\Lambda }=\frac{c^{2}}{H_{0}^{2}L^{2}}=\frac{c^{2}}{H^{2}L^{2}}%
E(x)^{2}.  \label{rhoL}
\end{equation}%
So
\begin{equation}
\tilde{\rho}_{\Lambda }^{^{\prime }}=\frac{d}{dx}\tilde{\rho}_{\Lambda }=(-2%
\frac{L^{^{\prime }}}{L})\tilde{\rho}_{\Lambda },  \label{rho-derivative}
\end{equation}%
\begin{equation}
E(x)\cdot E(x)^{\prime }=\frac{3}{2}[\tilde{\rho}_{m}+(1+w_{\Lambda })\tilde{%
\rho}_{\Lambda }+\frac{4}{3}\tilde{\rho}_{r}].  \label{Fredman-new2}
\end{equation}%
By substituting Eq.(\ref{rho-derivative}) into Eq.
(\ref{newrhoL}), we have
\begin{equation}
w_{\Lambda }(x)=-1-\frac{2}{3}\frac{L^{^{\prime }}(x)}{L(x)}-\frac{\lambda }{%
3}\frac{\tilde{\rho}_{m}(x)}{E(x)}.  \label{equation of state}
\end{equation}%
If $L$ is taken to be the particle horizon (Eq.(\ref{particle horizon})) or
future event horizon (Eq.(\ref{event horizon})), the derivative of $L$ is
\begin{equation}
L^{^{\prime }}=-(L\pm \frac{1}{H}),  \label{L derivative}
\end{equation}%
where the upper (lower) sign represents particle horizon (future event
horizon). By substituting Eq. (\ref{L derivative}) and (\ref{rhoL}) into Eq.
(\ref{equation of state}), we have the equation of state in the particle
horizon (upper sign) and future event horizon (lower sign) cases
\begin{equation}
w_{\Lambda }(x)=-\frac{1}{3}\pm \frac{2}{3}\frac{\tilde{\rho}_{\Lambda }^{%
\frac{1}{2}}}{c\cdot E(x)}-\frac{\lambda }{3}\frac{\tilde{\rho}%
_{m}(x)^{\alpha }\tilde{\rho}_{\Lambda }(x)^{\beta -1}}{E(x)}
\label{EoS for ph,eh}
\end{equation}
Substituting Eq. (\ref{EoS for ph,eh}) into (\ref{Fredman-new2}), we have
\begin{equation}
2E(x)\cdot E^{^{\prime }}(x)=3(\tilde{\rho}_{m}+\frac{4}{3}\tilde{\rho}%
_{r})+2\tilde{\rho}_{\Lambda }\pm 2\frac{\tilde{\rho}_{\Lambda }^{\frac{3}{2}%
}}{c\cdot E(x)}-\frac{\lambda }{E(x)}\tilde{\rho}_{m}^{\alpha }\tilde{\rho}%
_{\Lambda }^{\beta }.  \label{Intermediate1}
\end{equation}%
Taking the derivative of Eq. (\ref{Fredman-new1}) with respect to
$x$, we have
\begin{equation}
2E(x)\cdot E^{^{\prime }}(x)=\tilde{\rho}_{m}^{^{\prime }}+\tilde{\rho}%
_{\Lambda }^{^{\prime }}+4\tilde{\rho}_{r},  \label{Intermediate2}
\end{equation}%
Substituting Eq. (\ref{new rhom}) into Eq. (\ref{Intermediate2}), we obtain
\begin{equation}
2E(x)\cdot E^{^{\prime }}(x)=3\tilde{\rho}_{m}+\tilde{\rho}_{\Lambda
}^{^{\prime }}+4\tilde{\rho}_{r}-\frac{\lambda }{E(x)}\tilde{\rho}%
_{m}^{\alpha }\tilde{\rho}_{\Lambda }^{\beta }.  \label{Intermediate3}
\end{equation}%
Subtracting equations (\ref{Intermediate3}) from
(\ref{Intermediate3}), we obtain the following simple dynamical
equation:
\begin{equation}
\tilde{\rho}_{\Lambda }^{^{\prime }}=2\tilde{\rho}_{\Lambda }[1\pm \frac{%
\tilde{\rho}_{\Lambda }^{\frac{1}{2}}}{c\cdot E(x)}],  \label{motion3}
\end{equation}%
Eliminate $\tilde{\rho}_{m}$ in Eq. (\ref{Intermediate1}) using Eq.(\ref%
{Fredman-new1}),
\begin{equation}
E^{^{\prime }}(x)=\frac{3}{2}E(x)+\frac{1}{2E(x)}(\tilde{\rho}_{r}-\tilde{%
\rho}_{\Lambda })-\frac{\lambda }{2E(x)^{2}}(E(x)^{2}-\tilde{\rho}_{r}-%
\tilde{\rho}_{\Lambda })^{\alpha }\tilde{\rho}_{\Lambda }^{\beta }\pm \frac{%
\tilde{\rho}_{\Lambda }^{\frac{3}{2}}}{c\cdot E^{2}(x)},
\label{motion4}
\end{equation}%
and
\begin{equation}
w_{\Lambda }(x)=-\frac{1}{3}\pm \frac{2}{3}\frac{\tilde{\rho}_{\Lambda }^{%
\frac{1}{2}}}{c\cdot E(x)}-\frac{\lambda }{3}\frac{\tilde{\rho}_{\Lambda
}^{\beta -1}}{E(x)}(E(x)^{2}-\tilde{\rho}_{\Lambda }-\tilde{\rho}%
_{r})^{\alpha }.  \label{EEoS}
\end{equation}%
To make Eq. (\ref{motion3}) and (\ref{motion4}) easy to solve numerically,
we change the variable to $z$,
\begin{equation}
\frac{d\tilde{\rho}_{\Lambda }}{dz}=2\frac{\tilde{\rho}_{\Lambda }}{(1+z)}%
[1\pm \frac{\tilde{\rho}_{\Lambda }^{\frac{1}{2}}}{c\cdot E(z)}],\text{ }%
\tilde{\rho}_{\Lambda }(0)=1-\Omega _{m0},  \label{motion3'}
\end{equation}%
\begin{eqnarray}
\frac{dE(z)}{dz} &=&\frac{1}{(1+z)} [  \frac{3}{2}E(z)+\frac{1}{2E(z)}(\tilde{%
\rho}_{r}-\tilde{\rho}_{\Lambda })\pm \frac{\tilde{\rho}_{\Lambda }^{\frac{3%
}{2}}}{c\cdot E^{2}(z)}  \notag \\
&&-\frac{\lambda }{2E(z)^{2}}(E(z)^{2}-\tilde{\rho}_{r}-\tilde{\rho}%
_{\Lambda })^{\alpha }\tilde{\rho}_{\Lambda }^{\beta } ],  \notag \\
E(0) &=&1,  \label{motion4''}
\end{eqnarray}%
where
\begin{equation}
\tilde{\rho}_{r}=\Omega _{r0}(1+z)^{4}=4.78\times 10^{-5}(1+z)^{4},
\end{equation}%
and the upper (lower) sign in Eq. (\ref{motion3'}) and (\ref{motion4''})
corresponds to the model with particle (future event) horizon. Thus, Eq. (%
\ref{motion3'}) and (\ref{motion4''}) are the complete system of
equations. Eqs.(\ref{EEoS})-(\ref{motion4''}) reproduce
Eqs.~(\ref{event case}). In the limit $\lambda=0 $ and $\rho
_{r}=0,$ the equations reduce to the non-interacting holographic
dark energy model \cite{Li:2004rb,Wu07}. The evolution of event
horizon size $L$ and the corresponding dark energy density are
shown in Fig.~\ref{diverge}. As $z \to -1$, the size of the event
horizon shrinks to zero, and the dark energy density diverges,
i.e. we are approaching a Big Rip singularity.

\section{Big Rip Singularity}

There have been extensive discussions on the future singularity of the dark
energy models, particularly for the case where the future Universe is
dominated by the phantom dark energy. Ref. \cite{Elizalde04,Nojiri05a}
investigated the appearance of the future singularity, and showed that the
singularities could be classified into four types. The Big Rip is however, the
only relevant singularity in the holographic dark energy model \cite{Nojiri05a,
Elizalde05}. Ref \cite{Nojiri09}
showed that the coupling between dark matter and dark energy
may help to solve the coincidence problem, but not the Big Rip (Type \textrm{%
I}) singularity issue. In particular, \cite{Elizalde05} discussed the
singularity problem in the holographic dark energy model, and proved that:
\begin{enumerate}
\item if holographic dark energy $c=1,$ there exists a de Sitter
solution to the dynamic equation system of the HDE. But if $c\neq
1,$ the de Sitter solution does not exist. \item For the de Sitter
solution ($c=1$), one should consider the back reaction of the
quantum effects near the singularity. The contribution of the
conformal anomaly as a back reaction near singularity could indeed
moderate or even prevent the future singularity (Eqs.(53)-(61) in
\cite{Elizalde05}).
\end{enumerate}

In our interacting HDE model, the best-fit parameter $c=0.81$, and
for such model the fate of the Universe is a Big Rip in the
future. In the following, we investigate whether this conclusion
is affected by the back reaction of quantum effects.

Since the future of the Universe is dominated by dark energy, we
will neglect all of the terms which contain $\rho _{m}$ and $\rho
_{r}$, as they will become subdominant when compared with the dark
energy density $\rho _{\Lambda }.$ The dynamical equations then
simplify to
\begin{equation}
E(x)^{2}=\tilde{\rho}_{\Lambda },  \label{rho1_sim}
\end{equation}%
\begin{equation}
\tilde{\rho}_{\Lambda }^{\prime }-3(1+w_{\Lambda })\tilde{\rho}_{\Lambda }=0
\label{rho2_sim}
\end{equation}%
\begin{equation}
L^{\prime }=-(L-\frac{1}{H}),  \label{L_sim}
\end{equation}%
where we only consider the case for future event horizon. From Eqs. (\ref%
{rho1_sim})-(\ref{L_sim}), the asymptotical behavior of $H$,
$\tilde{\rho}_{\Lambda }$ and $w_{\Lambda }$ can be obtained easily:
\begin{equation}
w_{\Lambda }=-\frac{1}{3}-\frac{2}{3}\frac{1}{c},  \label{w_sim1}
\end{equation}%
\begin{equation}
\rho _{\Lambda }=\rho _{\Lambda 0}a^{2(\frac{1}{c}-1)},  \label{rho3_sim}
\end{equation}%
\begin{equation}
H=H_{0}\Omega _{\Lambda 0}^{\frac{1}{2}}a^{\frac{1}{c}-1},  \label{H_sim}
\end{equation}%
where $\rho _{\Lambda 0}=3H_{0}^{2}M_{pl}^{2}\Omega _{\Lambda 0},$ and $%
\Omega _{\Lambda 0}$ is the fractional HDE density taken at the present day.
From Eqs. (\ref{H_sim}), the derivative terms are
\begin{equation}
\dot{H}=\left( H_{0}\Omega _{\Lambda 0}^{\frac{1}{2}}\right) ^{2}\left(
\frac{1}{c}-1\right) a^{2(\frac{1}{c}-1)},
\end{equation}%
\begin{equation}
\ddot{H}=\left( H_{0}\Omega _{\Lambda 0}^{\frac{1}{2}}\right) ^{3}2\left(
\frac{1}{c}-1\right) ^{2}a^{3(\frac{1}{c}-1)},
\end{equation}%
\begin{equation}
\dddot{H}=\left( H_{0}\Omega _{\Lambda 0}^{\frac{1}{2}}\right) ^{4}6\left(
\frac{1}{c}-1\right) ^{3}a^{4(\frac{1}{c}-1)}.
\end{equation}%

The conformal anomaly $T_{A}$ (back reaction term) has the
following form \cite{Elizalde05}:
\begin{equation}
T_{A}=-12b\dot{H}^{2}+24b^{\prime }(-\dot{H}^{2}+H^{2}\dot{H}%
+H^{4})-(4b+6b^{\prime \prime })(\dddot{H}+7H\ddot{H}+4\dot{H}^{2}+12H^{2}%
\dot{H}).
\end{equation}%
Thus in the HDE model
\begin{equation}
T_{A}=12H_{0}^{4}\Omega _{\Lambda 0}^{2}f(b,c)a^{4(\frac{1}{c}-1)},
\label{ta}
\end{equation}%
where
\begin{eqnarray}
f(b,c) &=&2b^{\prime }\left[ -\left( \frac{1}{c}-1\right) ^{2}+\frac{1}{c}%
\right] -3b^{\prime \prime }\left[ \frac{1}{c}\left( \frac{1}{c}+1\right)
\left( \frac{1}{c}-1\right) \right]   \notag \\
&&-b\left( \frac{1}{c}-1\right) \left[ 2\left( \frac{1}{c}-1\right)
^{2}+7\left( \frac{1}{c}-1\right) +4\right] .  \label{fbc}
\end{eqnarray}%
With Eqs.(\ref{ta}) and (\ref{fbc}), the energy density and pressure of the
conformal anomaly $\rho_{A}$ are \cite{Elizalde05},
\begin{eqnarray}
\rho _{A} &=&-\frac{1}{a^{4}}\int_{t_{0}}^{t}dta^{4}HT_{A}
\simeq -3H_{0}^{4}\Omega _{\Lambda 0}^{2}cf(b,c)a^{4(\frac{1}{c}-1)},\\
p_{A} &=&\frac{1}{3}\left( T_{A}+\rho _{A}\right)
=4\left( 1-\frac{c}{4}\right) H_{0}^{4}\Omega _{\Lambda 0}^{2}f(b,c)a^{4(%
\frac{1}{c}-1)}.
\label{rhoA}
\end{eqnarray}%
Under this back reaction, the expansion rate is
\begin{eqnarray}
H^{2} &=&\frac{1}{3M_{pl}^{2}}\left( \rho _{\Lambda }+\rho _{A}\right)
\notag \\
&=&H_{0}^{2}a^{2(\frac{1}{c}-1)}\left[ \Omega _{\Lambda 0}-\left( \frac{H_{0}%
}{M_{pl}}\right) ^{2}\Omega _{\Lambda 0}^{2}f(b,c)a^{2(\frac{1}{c}-1)}\right]
.  \label{hubble2}
\end{eqnarray}%
Since the energy density of back reaction is positive ($\rho
_{A}>0)$, the function $f(b,c)<0;$ therefore, $H^{2}$ increase
monotonically in the future if $c<1.$ Thus, we prove that for the
$c<1$ case, the back reaction effect from quantum corrections
cannot prohibit the occurrence of the Big Rip.

\end{document}